  \documentclass{aastex}
  \usepackage{emulateapj5}

\newcommand{\Bvec}{{\bf b}}
\newcommand{\uvec}{{\bf u}}
\newcommand{\los}{{line-of-sight}}

\newcommand{\reyn}{{\hbox{Re}}}
\newcommand{\reynm}{{\reyn_m}}

	\newcommand{\citeN}[1]{\citeauthor{#1} (\citeyear{#1})}
	\newcommand{\citeNP}[1]{\citeauthor{#1} \citeyear{#1}}

\shorttitle{Polarization of lines from turbulent dynamo simulations}

\shortauthors{S\'anchez Almeida, Emonet \& Cattaneo}

\begin{document}


\title{Polarization of photospheric lines from turbulent dynamo
simulations}


\author{J. S\'anchez Almeida}
\affil{Instituto de Astrof{\'{\i}}sica de Canarias, 
E-38200 La Laguna, Tenerife, Spain}
\email{jos@ll.iac.es}

\author{T. Emonet}
\affil{The University of Chicago, Department of Astronomy and
Astrophysics, 5640 South Ellis, Chicago, IL 60637, USA}
\email{emonet@flash.uchicago.edu}

\and

\author{F. Cattaneo}
\affil{The University of Chicago, Department of Mathematics, 5734
S. University, Chicago, IL 60637, USA} 
\email{cattaneo@flash.uchicago.edu}


\begin{abstract}
We employ the magnetic and velocity fields from  turbulent
dynamo simulations 
to synthesize the polarization of a typical photospheric
line.
The synthetic Stokes profiles have properties in common with
those observed in the quiet Sun. The simulated magnetograms present a
level of signal similar to that of the Inter-Network regions. Asymmetric
Stokes $V$ profiles with two, three and more lobes appear in a natural
way. The intensity profiles are broadened by the magnetic fields in
fair agreement with observational limits. Furthermore, the Hanle
depolarization signals of the \ion{Sr}{1}~$\lambda$4607~\AA\ line turn out to
be within the solar values.
Differences between synthetic and
observed polarized spectra can also be found.
There is a shortage of Stokes $V$ asymmetries, that
we attribute to a deficit of structuring in the 
magnetic and velocity fields from the simulations as compared to the Sun. 
This deficit may reflect the fact that the Reynolds numbers of the
numerical data are still far from solar values.
We consider the possibility that
intense and tangled magnetic fields, like those 
in the simulations, exist in the Sun. This scenario
has several important consequences. For example,
less than 10\% of the existing unsigned magnetic flux would be detected 
in present magnetograms. The existing flux would exceed by far
that carried by active regions during
the maximum of the solar cycle. Detecting these magnetic fields 
would involve improving the angular resolution,
the techniques to interpret the polarization signals, and to a
less extent, the polarimetric sensitivity.
\end{abstract}




\section{Introduction}\label{sec_introduction}

\begin{figure}
\epsscale{.8}
 \plotone{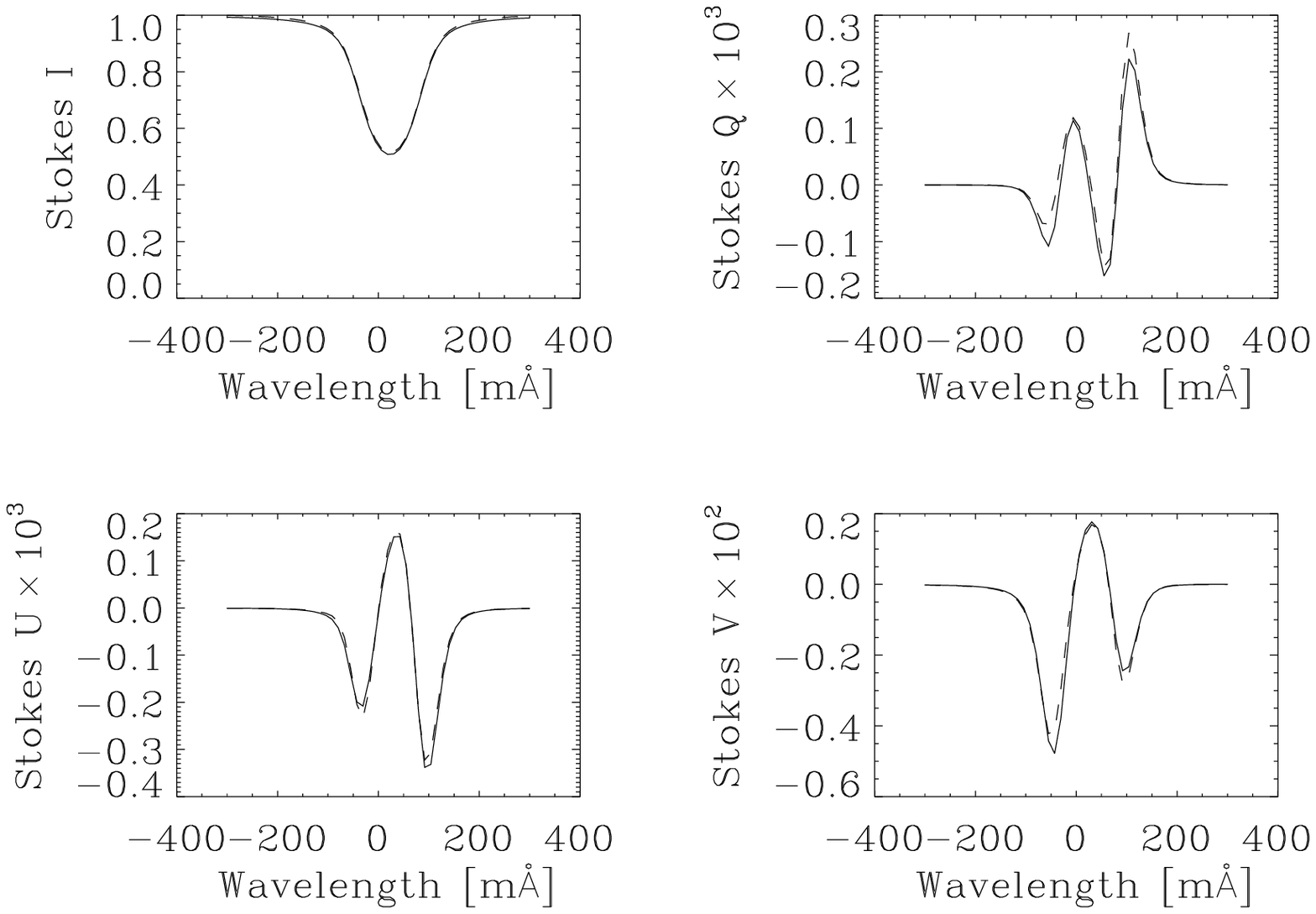}
 \caption{Approximation used to synthesize the
	polarized spectrum. The solid lines show Stokes 
	$I$, $Q$, $U$ and $V$ profiles
	from a point of the simulation synthesized according to the
	MISMA approximation employed along the text. The dashed lines
	represent a full integration of the radiative transfer equations.
	Typical relative deviations are of the order of a few per cent.
	Wavelengths are in m\AA\ and the Stokes profiles have been normalized to
	the continuum intensity.}
 \label{full_vs_misma}
 \end{figure}

In the absence of gradients of velocity and magnetic field
in the resolution element,
the polarization emerging from an atmosphere must be
either symmetric or antisymmetric with respect to the
central wavelength  of each spectral line (e.g., \citeNP{unn56};
\citeNP{lan83b}). 
The polarization observed in the solar photosphere 
does not show such symmetries.
Asymmetric Stokes profiles\footnote{
The term {\em Stokes profile} denotes the variation within
a spectral line of any of the four Stokes parameters. Using the
standard nomenclature, we use Stokes $I$ to represent
the intensity, Stokes $Q$ and $U$ for the two independent
types of linear polarization, and Stokes $V$ for the degree
of circular polarization.
Examples are given in Figure  \ref{full_vs_misma}.
}
arise from the quiet Sun (e.g., \citeNP{san96}; \citeNP{sig99}), 
from plages and  network regions (e.g., \citeNP{bau81}; \citeNP{ste84}), 
as well as from sunspot penumbrae 
(e.g., \citeNP{gri72}; \citeNP{mak86}; \citeNP{san92b}). 
The fact that such asymmetries are found even when observing at the limit of
present resolution suggests
that a rich structuring remains 
unresolved to the current observations. 

Because asymmetries carry information on the spatially unresolved
properties of the 
photospheric plasma, 
their study and correct interpretation offers a chance to  
overcome the limitations imposed by the angular resolution, and to
retrieve information inaccessible to direct imaging.
Indeed, such possibility has been exploited during the last decade,
always relying on a considerable amount of modeling and assumptions.
In broad terms, one can distinguish two approaches depending on the 
size of the unresolved structures that are responsible for the asymmetries.

The first approach assumes that the unresolved
photospheric structures are actually on the verge  of being
resolved in broad-band images taken with the current instrumentation. 
The smallest detectable features are of the order of 0.2\farcs  (or 150 km on the Sun);
a size set by technical limitations of the present
solar telescopes (e.g., \citeNP{bon99}).  
Several models  of this kind
have been proposed to explain asymmetries in special cases
like penumbrae, plage and network regions, and in the quiet Sun 
(e.g., \citeNP{sol93b};
	\citeNP{gro88}; \citeNP{bel97};
	\citeNP{stei00}). For 
example, \citeN{stei00} 
invokes special thermal structures (temperature inversion)
within magnetic regions in order to explain  some
extreme Stokes $V$ shapes frequently observed
in the quiet Sun. The distinctive feature of this approach is that the
proposed configurations (thermal, magnetic or kinetic) are 
specific to the case under study. The fact that asymmetries occur everywhere is
therefore difficult to explain within this framework.

The second approach is more consonant
with the ubiquity of the asymmetries in the polarization of
photospheric lines. It assumes that the
photospheric plasma is in a turbulent state and that small-scale structures both in velocity and
magnetic field are present. Order of magnitude estimates for the magnetic and
kinetic Reynolds numbers in the granulation indicate that
the magnetic field could be structured on spatial scales as small as a
few kilometers (e.g., \citeNP{sch86}). 
This picture is consistent with recent progress in dynamo theory suggesting that a substantial part of 
the magnetic
field in the quiet Sun could be generated locally by dynamo action driven by the
granular flows (\citeNP{men89}; \citeNP{pet93}; \citeNP{cat99a}; \citeNP{emo01}).  
This second approach is  also
supported by certain  observations indicating that the magnetic field 
is structured on spatial scales below the resolution
limit of current telescopes (e.g., \citeNP{ber96}; \citeNP{ber98};
\citeNP{san98a}).
Based on these premises, it is reasonable to think of the asymmetries 
as the result of 
observing  magnetic fields that vary spatially on
scales much smaller than the
mean-free-path of the photons.
Interpretations of Stokes profiles  taking
into account this very fine structuring of the atmosphere have been
carried out for several years (\citeNP{san96}).
The model atmospheres used to fit the observed
profiles consist of a collection of magnetic and non-magnetic
components, each containing mild gradients
to comply with the height variations of the mean photosphere,
but interleaved in such a way as to produce large
gradients along the \los\ (\citeNP{san97b}).
The appealing aspect of this approach is that asymmetries emerge spontaneously, and
independently of the details of the model; 
arising from correlations between the magnetic and the velocity fields
of the various components (e.g., asymmetries occur if, on average, downflowing elements have low magnetic field
strength). 
It is known that
atmospheres with micro-structure reproduce all kinds of
Stokes profiles observed in the quiet Sun, including network and 
Inter-Network (IN) regions (\citeNP{san00}; \citeNP{soc02}). 
These MIcro-Structured Magnetic Atmospheres (MISMAs) portray 
a quiet Sun with large amounts of unsigned magnetic flux and 
very complex magnetic topology (very often two magnetic
polarities coincide in a resolution element).

In the present paper we take advantage of recent developments in the
numerical modeling of surface dynamos to understand the origin
of asymmetries in the line polarization. 
We use the magnetic and velocity fields  from a set
of numerical data generated  to study the interaction between thermally driven turbulent convection and magnetic
fields (\citeNP{cat99a}; 
\citeNP{emo01}).
Although these numerical simulations were not designed for a detailed
description of the solar photosphere, 
the complexity and ubiquity of its fields recall
in many respects the magnetic quiet Sun inferred from
the observed asymmetries. 
Assuming a
Milne-Eddington (ME) atmosphere for the thermodynamic variables we produce
synthetic Stokes profiles. Thus, 
the asymmetries in the
resulting profiles are directly related to the correlations between
the velocity and the magnetic field that exist in the numerical data, but they
are decoupled from the thermodynamic variables of the simulation.
The comparison between the synthetic profiles and solar data
shows that the synthetic spectra are frequently similar
to the observed ones. Such agreement suggests that the simulation 
includes some of the ingredients that characterize the quiet
Sun magnetic fields; in particular, the correlations between
magnetic field and velocity at the smallest spatial scales.
On the other hand, discrepancies between synthetic and observed Stokes profiles allow to identify 
missing ingredients and, consequently, to devise strategies to improve 
the simulations and the inversions.
Finally, following \citeN{emo01}, one can use the
synthetic profiles to 
estimate the angular resolution required to
determine the basic properties of the magnetic structures present in
the numerical simulation. This angular resolution may be of relevance
to decide the specifications of new solar instruments.

The work is organized as follows. The numerical data
is briefly described in \S \ref{sec_mhd}. 
The synthesis procedure is detailed in \S \ref{sec_spectra} and  
includes two subsections: in \S \ref{seeing} 
we characterize the angular resolution of the model observations, 
and in \S \ref{calibration} we calibrate the synthetic magnetograms.
The main results of the synthesis are 
analyzed in \S \ref{sec_results}:
the variation of the apparent flux of the region depending on the sensitivity
and spatial
resolution of the observation (\S \ref{flux_density}),
the asymmetries of the Stokes $V$ profiles (\S \ref{asym}),
and the magnetic broadening of the intensity profiles (\S \ref{broadening}).
Hanle depolarization signals to be expected 
for \ion{Sr}{1} $\lambda$4607~\AA\ are worked out in \S \ref{hanle}.
\S \ref{telescope} discusses the diameter of the ideal telescope
needed to spatially resolve the simulations.
Finally, the implications of the present work are discussed
in \S \ref{conclusions}.

\section{Description of the set of MHD data}\label{sec_mhd}
The numerical data set used in the present work is part of a series of
numerical simulations  to study the generation and interaction  of magnetic
fields with turbulent convection. The simulations are based on a idealized model 
describing a layer of incompressible (Boussinesq) fluid with constant kinematic viscosity  $\nu$,
thermal diffusivity $\kappa$ and magnetic diffusivity $\eta$. The
boundary conditions are periodic in the horizontal directions, and impenetrable and
stress free with constant temperature and zero horizontal magnetic
field along the upper and lower boundaries of the computational domain. 
As is common in this kind of simulations, the unit of length is the vertical extent of the layer, the unit of time is
the thermal diffusion time
across that layer and the magnetic intensity is expressed as an
equivalent Alfv\'en speed.  With these units the 
particular dynamo solutions used below is defined by the following dimensionless parameters: aspect-ratio of the
computational domain
$10\times 10 \times 1$, Rayleigh number $5\times 10^5$ and Prandtl
numbers $\nu/\kappa=1$, $\nu/\eta=5$. The numerical resolution 
was $512\times 512\times 97$ collocation points. 
In this regime, the solution corresponds to a state of vigorous convective turbulence characterized by a  
kinetic and magnetic Reynolds numbers of  
$\reyn=200$  $\reynm\simeq
1000$, respectively. The resulting flow is strongly chaotic and acts as an efficient dynamo generating
an intermittent magnetic field with no mean flux and a magnetic energy roughly  20\% of the kinetic energy.
For the present study, we use one time step of the dynamo evolution
that is well into the statistically stationary regime. At this epoch,
the  rms velocity of the fluid  is $u\approx 200$ 
with a corresponding turnover time\footnote{We define the turnover
time to be twice the vertical crossing time based on the rms speed
$u$.} of $1/100$. 
A more complete description of the simulation procedure, and further details about the numerical solutions can be
found in \citeN{cat99a}, \citeN{emo01} and \citeN{cat01}. 

\section{Spectral synthesis}\label{sec_spectra}

The numerical simulations described above were not designed specifically for spectral
synthesis, rather to study dynamo action. In order to achieve the high magnetic Reynolds numbers ($\reynm=1000$)
necessary for dynamo action to manifest itself, and given finite computer resources ($10^7$--$10^8$ grid
points), we found it necessary to resort to the Boussinesq approximation. The latter is valid provided that the
vertical extent of the layer is much smaller than  the pressure, density, and temperature scale heights (see
e.g. \citeNP{spi60}). As a consequence, the thermodynamic variables
from the numerical solutions have modest variations across the layer, and cannot be used to calculate the opacity
and source functions needed for the synthesis. Consequently, we address the
synthesis problem ignoring the thermodynamic variables from the
simulations, but keeping the variations in the velocity ($\uvec$)  and magnetic field ($\Bvec$). 
Although $\uvec$ and $\Bvec$ are then decoupled from the density and
temperature distributions, the resulting synthetic profiles are useful to study the properties 
of the emerging line polarization  that depends, primarily, on the 
magnetic and velocity  structure.
In the remainder of this section we explain the process followed to 
generate the synthetic profiles.

The first step is to translate the dimensionless velocity and magnetic data
of the simulations into units that
are practical for the synthesis. The spatial scale is fixed by
assuming that the typical size of one granule is about 1000 km. In the
statistically stationary regime the computational domain is about 5 granules wide. Thus
the horizontal  size of the domain corresponds to 5000 km, with a corresponding grid resolution of approximately
$10$~km in the horizontal, and $5$~km in the vertical.
Magnetic field strengths $B$ and velocities $U$ readily follow from
the energy equipartition arguments.
$B$ and $U$ are proportional to
the dimensionless 
magnetic strength $b$ and velocity $u$
through the factors $f_u$ and $f_b$,
\begin{eqnarray}
B=&f_b b,\qquad
U=&f_u u.\label{scl0}
\end{eqnarray}
Equipartition between kinetic energy density and magnetic
energy density is reached in dimensionless units when
\begin{equation}
b_{eq}^2=u_{eq}^2,
\label{scl1}
\end{equation}
or equivalently when
\begin{equation}
B_{eq}^2/(8\pi)=\rho U_{eq}^2/2\;,
\end{equation}
$\rho$ being the density. The previous equation,
together
with equations (\ref{scl0}) and (\ref{scl1}), lead to
\begin{equation}
f_b=f_u (4\pi\rho)^{1/2}\;.
\label{scl2}
\end{equation}
Thus, if one associates the rms speed $u\simeq 200$ with the rms
velocity fluctuations observed in the solar granulation $U=2$ km
s$^{-1}$ (e.g., \citeNP{dur82}; \citeNP{top91}), then
\begin{equation}
f_u= 10^{-2}\ {\rm km\ s}^{-1}\;,
\label{scl3}
\end{equation}
and for a typical density at the base
of the photosphere (3 $\times 10^{-7}$ g cm$^{-3}$; \citeNP{mal86}),
equations (\ref{scl2}) and (\ref{scl3}) render
\begin{eqnarray}
f_b\simeq 2\ {\rm G}\;.
\label{scl4}
\end{eqnarray}
Except for a few exploratory calculations described in \S \ref{asym} and
\S \ref{hanle},
the scaling factors in equations (\ref{scl3}) and (\ref{scl4}) are
used for the rest of the work. Notice that reasonable deviations from this
scaling do no affect our results in a substantial way (a point addressed in \S
\ref{sec_results}).

Following our previous assumption,
the absorption and emission terms of the radiative
transfer equations (e.g., \citeNP{bec69a}; \citeNP{wit74};
\citeNP{lan76}) are not based on the thermodynamic variables of the
numerical data set. Instead, we resort to a Milne-Eddington (ME) synthesis
which assumes a constant line absorption and a linear
source function (e.g., \citeNP{unn56}; \citeNP{lan92}).
The ME assumption is routinely used for magnetic field diagnostics
(e.g., \citeNP{soc01}) since it yields
an analytic expression for the emerging
polarized spectrum that only depends on a few parameters.
The analytic solution is used to fit observed spectra and retrieve
information from them (e.g., \citeNP{sku87}).
The ME synthesis hides
all the unknown thermodynamic properties
of the atmosphere in  
five parameters, namely, the line absorption coefficient,
two parameters that define the (linear) source function, the damping
coefficient, and the Doppler width. In our syntheses, we adopt the
values characteristic of the  
\ion{Fe}{1} line at $\lambda$6302.5 \AA\ in the
quiet Sun (\citeNP{san96}, Table 1, including a Doppler width of
40 m\AA\ or 1.9 km s$^{-1}$). This magnetically sensitive
spectral line is often used for magnetic studies, including those of the Advanced
Stokes Polarimeter (\citeNP{elm92}). Working with the same line
allows us to compare our results directly with its observations.
The use of observed ME thermodynamic parameters to represent
the synthetic line provides realism to 
the part of the synthesis that we do not obtain 
directly from the numerical simulation.
The emerging spectra will have the observed
equivalent widths, line widths, 
etc\footnote{Two independent results of the synthesis reflect this realism.
First, the constant to calibrate magnetograms worked out
in  \S \ref{calibration} agrees with those derived from real solar spectra.
Second, the mean equivalent width of our synthetic intensity profiles is
71 m\AA , i.e., close to the value observed in the quiet Sun
(some~80~m\AA; e.g., \citeNP{moo66}).}.

The standard ME assumption considers an atmosphere having
a uniform magnetic field  (e.g., \citeNP{unn56}; \citeNP{lan92}). 
Such assumption is clearly at odds  with our numerical data
in which the magnetic diffusive scale
is of the order of $500\times\reynm^{-1/2}$~km $\simeq 15$~km,
i.e., three grid points in the vertical direction.
The field is therefore not constant over the range of heights where 
the typical photospheric lines are formed (say 100 km). 
A realistic spectral line synthesis would have to take into account the
contributions from many layers in the simulation. 
The  brute-force approach would be to assign an optical
depth to each horizontal plane in the simulation
and then synthesizing the
spectrum by direct numerical integration of the radiative
transfer equations for polarized light. Here, instead, we resort to
a different strategy that a) is simpler and much faster, 
b) reduces the number of free parameters of the synthesis (no
need to specify the optical depths of all the different planes
of the atmosphere) and, c) produces spectra
very close to those obtained by the brute-force approach.
We employ a multi-component
ME atmosphere, as defined by \citeauthor{san96} (\citeyear{san96}; \S 2.1).
We assume that not all individual planes of the simulation contribute to the emerging spectrum, and consider only a
sub-set whose thickness is of the order of the photon mean-free-path.
We take into account  that the planes  are optically
thin so that the emerging spectrum depends only  on the average along the \los\ of the emission 
and the absorption  (the so-called MISMA approximation;
\citeNP{san96}).
As argued below, these
assumptions are not unreasonable,
and  reduce the radiative
transfer calculation 
to deciding how many planes contribute to the
emerging spectrum. Subsequently,  the
average absorption matrix is computed, and used jointly with the
analytic solution for the regular ME synthesis
(\citeNP{san96}). 

In our case we choose the 
21 uppermost planes of the simulation, excluding the upper boundary.
According to the scaling described in the previous paragraph,
this corresponds to approximately 100 km in the atmosphere, a typical range
for the formation of a photospheric line. 
We further assume that the numerical simulation is  placed at disk center, i.e., with the \los\ 
along the vertical direction.

Since the individual planes of the simulation are indeed optically thin (5 km),
using the average emission and absorption 
instead of solving the full radiative transfer
should be a good approximation. We checked this assumption by 
comparing spectra computed by direct integration of the
radiative transfer equation with those resulting from our approximation.
Figure \ref{full_vs_misma} contains  one example.
The full synthesis is performed assuming the 21 different planes
of the ME synthesis to be equi-spaced in  and spanning from
$\log\tau=0$ to $\log\tau=-1$. 
(The symbol $\tau$ stands for the 
continuum optical depth.)
This layer is periodically 
repeated to complete the atmosphere from $\log\tau=10$ to $\log\tau =-3$.
The synthesis of 1000 randomly chosen spectra
shows relative differences between the two kinds of syntheses
of only a few per cent. This deviation is negligible for the 
analysis that we carry out in the present work.  The choice of number of planes to be used for the 
spectral synthesis is a more delicate matter. Using 26 instead of 21 planes 
we find differences between spectra of up to 25\%. These
differences are due to the strong variability of the magnetic 
conditions, an uncertainty that also affects any other way of synthesizing
the spectra. For example, in the case of the brute-force approach
the signals depend on  the optical depths arbitrarily
assigned to each plane of the numerical data.

\subsection{Spatial resolution: seeing and telescope diameter \label{seeing}}

Studying the consequences of observing the simulations with a finite
spatial resolution is one of the key objectives  of the present work.  
We use the term {\em seeing} to  denote all the effects that may
reduce the spatial resolution (from genuine seeing to optical aberrations
of the instruments).
Here the effects of seeing are modeled by smearing 
the 2D maps of the Stokes profiles with Gaussian functions.
Note that the 2D convolution is carried
out at each individual wavelength of each Stokes parameter.
The amount of smearing is characterized by the 
FWHM (Full Width Half Maximum) of the Gaussian function.
We take this particular point spread function for the sake of simplicity, however,
it accurately describes 
long-integration-time atmospheric seeing 
(e.g, \citeNP{rod81}, \S 4.5).
In one of the sections  (\S \ref{telescope}),
we measure seeing  in terms of the diameter, $D$,
of an ideal telescope that  provides a given angular resolution.
Considering that the Airy disk of an ideal
telescope has a FWHM $\simeq\lambda/D$, then
\begin{equation}
	D=13\ {\rm cm\  (\lambda /6302)\ (FWHM/725)^{-1}},
	\label{eq1}
\end{equation}
where FWHM is expressed in km and $\lambda$ in \AA .
We have used a scale of 725 km/\arcsec , corresponding
to our target at 1 AU.

\subsection{Longitudinal Magnetograms \label{calibration}}

Longitudinal  magnetograms are images showing 
the degree of circular polarization
in the flank of a spectral line (e.g., \citeNP{bab53};
\citeNP{lan92}; \citeNP{kel94}).
They can be produced using relatively simple and compact instruments and
therefore have been widely used in solar magnetic field studies.
A significant fraction of what we know about quiet Sun 
magnetic fields has been derived from them (e.g., \citeNP{wan95}, and
references therein). 
Longitudinal magnetograms are usually calibrated in units of magnetic field strength
(G). Since we wish to compare the  polarization signals 
with observed magnetograms, calibrated magnetograms must be
prepared out of the synthetic Stokes profiles. 
We follow a procedure that mimics the observational process.
First, the circular polarization
signals are averaged in wavelength, this is
because magnetograms are often obtained through a rather broad
color filter. We employ a running box filter, 100 m\AA\ wide, centered in the
blue wing of the line at
50 m\AA\ from the line center. (These values are comparable to those
employed in real measurements; e.g., \citeNP{yi93b};
\citeNP{ber01}.)
Then the wavelength averaged circular polarization signal
$V_{a}$ is calibrated
to yield the so-called longitudinal magnetic flux
density $B_{\rm los}$,
i.e., 
\begin{equation}
B_{\rm los}=C_{\rm cal}\thinspace V_a.
	\label{bruteforce}
\end{equation}
The calibration constant $C_{\rm cal}$
is evaluated using the magnetograph equation and the properties
of the line in which the magnetograph operates;  explicitly,
\begin{equation}
C_{\rm cal}^{-1}=-c\lambda_0^2 g_{\rm eff} {{d I_a}\over{d\lambda}}.
\end{equation}
Here, the symbol $\lambda_0$ stands for the central
wavelength of the line, $g_{eff}$ is the
effective Land\'e factor, and
$I_a$ corresponds to the intensity profile smeared with the same color filter used to
produce $V_a$. The constant $c$ equals 4.67~$\times~10^{-13}$~\AA$^{-1}$~G$^{-1}$.
For the atomic parameters of \ion{Fe}{1} $\lambda$6302.5 \AA , and
using the mean Stokes $I$ profile over a complete snapshot,
we find
$C_{\rm cal}\simeq~5850$~G when $V_a$ is given in units of 
the continuum intensity.
This  synthetic  calibration constant is in good agreement with
the values that can be found in the literature for this line
(e.g., \citeNP{yi93b}; \citeNP{lit94}; \citeNP{soc02}).

With  several additional simplifying assumptions\footnote{Weak magnetic field that is
constant along the \los ,
Stokes $I$ independent of the magnetic field,  and others; see, e.g.,
\citeN{lan73}; \citeN{jef91}.
},
it can be shown that
\begin{equation}
B_{\rm los}=\int_S B_zdS \Big/\int_S dS,
	\label{theory}
\end{equation}
where the integrals extend to the resolution
element $S$, and $B_z$ is the
component of the magnetic field vector pointing towards the observer.
This identity provides the rationale to denote
magnetogram signals as {\em longitudinal magnetic flux densities}.
(According to equation [\ref{theory}], $B_{\rm los}$ 
represents the magnetic flux per unit area.)

\section{Results} \label{sec_results}

In this section we analyze the synthesis of the  512 $\times$ 512
Stokes spectra calculated from 
the dynamo simulation described in \S \ref{sec_mhd}.
We only use the numerical solution at one instant  since the purpose of the
work is to study the kind 
of spectra  characterizing the stationary state of the
simulation (studies of the time dependent behavior
is deferred for later).
We place the simulation at the solar disk center so that the
vertical direction follows the \los . 

\begin{figure}
\epsscale{.8}
\plotone{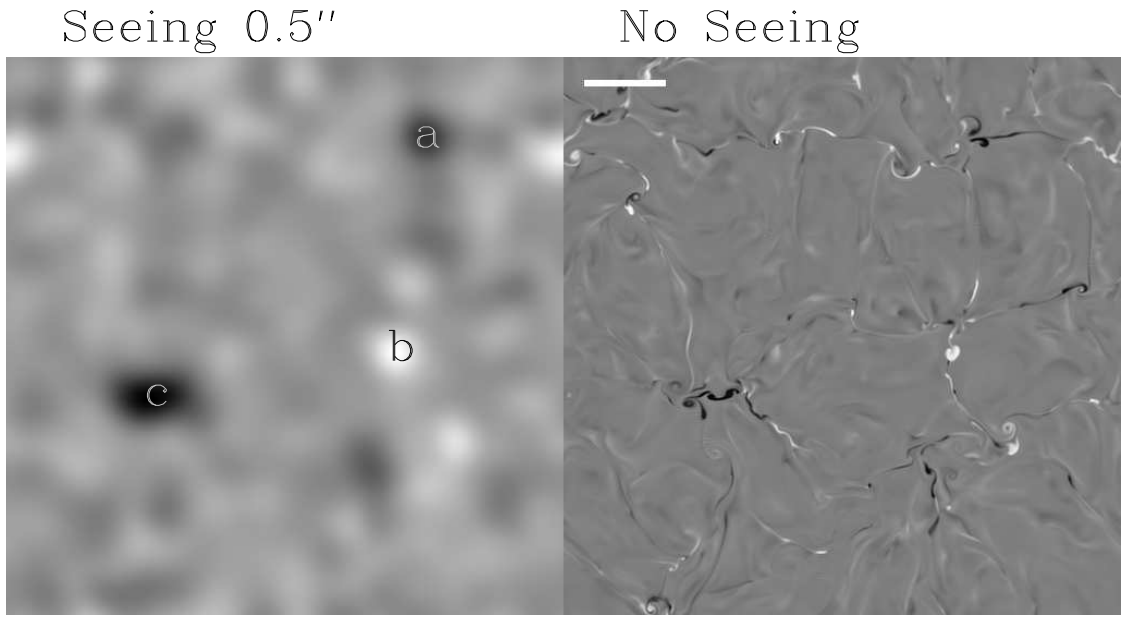}%
\caption{Magnetogram of the snapshot of the numerical simulation
analyzed in the work.
It shows both the original one at full resolution (right), as well as  the result of 
smearing the magnetogram with a 0\farcs5 seeing (left). The
white gauge indicates 1\arcsec\ on the Sun. The labels $a$, $b$ and
$c$ point out three entities that may be identified as 
single magnetic concentrations in the 0\farcs5 seeing magnetogram.
The scale of grays of the two magnetograms is independent since
the seeing diluted signals (left) would remain almost unnoticed 
if scaled to those 
in the original magnetogram (right).
}
\label{magneto}
\end{figure}

Figure \ref{magneto} shows the magnetogram that results when applying
equation (\ref{bruteforce}) to the synthetic Stokes $V$ spectra (right panel).
The effect of observing with a 0\farcs5 seeing -- representative of the best
angular resolution achieved at present (e.g., \citeNP{ber01}), is also
shown on the left panel. 
Ignoring the underlying substructure, an  observer would identify a number of magnetic
concentrations in the smeared magnetogram (e.g., points $a$,
$b$ and $c$). However, these are  difficult to associate with single structures   
once the underlying
substructure is acknowledged. 
Rather, the points with enhanced signal
in the 0\farcs 5 magnetogram represent locations where
the convective flows are continuously
advecting magnetized plasma to balance the plasma that constantly
disappears in the downdrafts 
(see also \S~3.2 in \citeNP{emo01}). The fact that 
the residual polarization signal shows up 
suggests the presence of a large-scale
structure of the advecting velocity. Indeed, detailed analysis of the numerical
data reveals the existence of a mesogranular flow (\citeNP{cat01}).

Except for a few test calculations in \S \ref{asym}, the spectra
discussed in this section correspond to a scaling of the dimensionless
magnetic field and velocity given by equations (\ref{scl3}) and (\ref{scl4}). 
However, syntheses using $f_u=0.02$ km s$^{-1}$ and $f_b=2.5$ G were also tried.
We found that
the degree of circular polarization increases linearly as $f_b$, whereas a
larger $f_u$ enhances the line asymmetries. On the other hand, the kind of 
general trends and properties that we analyze here do not depend
on the precise value of the scale factors.

\subsection{Flux density versus angular resolution \label{flux_density}}

The amount of magnetic flux and energy in
the numerical data is far larger than that detected in the Sun
as IN fields.
This difference can be understood as the result of two important
factors that hinder the detection of the
weak signals emerging from the simulation.
First, the magnetic fields are highly
disorganized so that the
polarization signals tend to cancel
as the angular resolution deteriorate (\citeNP{emo01}). Second, 
most synthetic  signals are extremely low, i.e.
at and below the sensitivity of the present 
instrumentation. Consequently a large fraction could not be detected at present.
In this section we use our synthetic lines to explore these effects.
We find that once realistic angular resolution
and sensitivity are taken into account, the simulations are in
good agreement with the observations.

Figure {\ref{mean}} shows the mean unsigned signal 
in the magnetogram of Figure \ref{magneto}
as a function of the seeing (i.e., mean $|B_{\rm los}|$ over the snapshot
versus FWHM of seeing). We consider
various sensitivities of the magnetograph:
moderate (20 G), good (5 G) and very good (0.5 G).
We account for the limited sensitivity by
setting to zero all those points in the
smeared magnetogram where the signal is
below the hypothetical observational threshold. 
Figures \ref{mean}b is identical to Figure \ref{mean}a
except that it has been normalized to the mean 
longitudinal field in the simulation 
(i.e., mean $|B_z| \simeq 51$~{\rm G}, 
where the average considers all the points of the
simulation that we use). 
According to the standard interpretation (equation [\ref{theory}]),
this is the parameter that one retrieves
from a magnetogram.
The normalization helps visualizing the fraction of real signal that
remains in the magnetogram for a given angular resolution
and sensitivity. There are several features in these
two figures that deserve comment.

Even with perfect sensitivity and maximum angular resolution, 
one detects only 80~\% of the existing flux.
The cancellation is mostly due to the averaging along the line-of-sight
caused by the radiative transfer.
At maximum angular resolution, the flux in the magnetogram
depends little on the sensitivity since most of the signals
exceed 20~G.

The decrease in signal strength as
the angular resolution deteriorate is severe. For the typical 1\arcsec\
angular resolution, the detectable signals are only 10 \% of the original 
ones (Fig. \ref{mean}b). This estimate is optimistic since it holds true
if the sensitivity
is good; should the latter be moderate, only traces of the original signals remain (1\% for 
20 G sensitivity).

Figure \ref{mean}a includes values for solar IN magnetic flux densities
observed by various authors.  
The level of
detected flux density agrees
well with the predictions of the simulations once the 
angular resolution has been taken into account.
Note that the observed flux densities are more than
a factor of ten smaller than the signals in the original magnetogram.
We should not overemphasize the agreement, since the
observational points are rather uncertain (they come from
inhomogeneous sources, with different sensitivities and
based on disparate techniques: see below). 
However, two conclusions can be drawn.
First, intense yet tangled magnetic fields like those in the
numerical simulation are compatible with the present
solar observations. Second, if our spatially fully resolved synthetic
spectra were emitted by the Sun, they would produce 
the degree of polarization that we detect on the Sun
with the present instrumentation.

The observations presented in Figure \ref{mean}a
required transformation from the numbers  quoted
in the original works to the 
quantities  plotted in the figure.
Here we briefly describe these transformations for the 
sake of completeness. (The symbols accompanying the citations correspond
to those used in Fig. \ref{mean}a.)
\citeauthor{soc02}~(\citeyear{soc02}; bullet sign $\bullet$) deduce 10 G
mean flux density for the quiet Sun fields, however, they mention that the
apparent flux density decreases by a factor 2.4 if 
it is estimated from magnetograms. This renders 4.2 G for a
1\arcsec~angular resolution, a figure derived
from the cut off frequency in the Fourier domain
of the continuum intensity image (\citeNP{san00}).
The same
procedure is used to estimate the angular resolution
of the spectra in 
\citeauthor{col01}~(\citeyear{col01}; square $\Box$).
The mean flux density, 3.4 G, has been directly
provided by the author. 
\citeauthor{wan95}~(\citeyear{wan95}; asterisk $\ast$)
point out, explicitly, a 1.65 G flux density for 
their angular resolution of 2\arcsec.
\citeauthor{sto00}~(\citeyear{sto00}; times sign $\times$)
find some 3.5 G mean flux density (private
communication), to which we associate an angular resolution
of two pixels or 1\farcs 4.
\citeauthor{lin99}~(\citeyear{lin99}; triangle $\bigtriangleup$)
do not directly give a mean flux density for their measurements.
However, their Figure 3  contains the magnetic fluxes of individual
measurements, and the authors provide the scale factor
between the magnetic flux and the magnetic flux density 
($5 \times 10^{15}$
Mx G$^{-1}$). With this conversion, the mean flux density of these
observations is about
4 G. The  magnetic features 
occupy 68 \% of the surface, since the rest remains
below the sensitivity threshold. Consequently, 
the mean flux density over the whole
surface results 0.68~$\times~4$~G. 
\citeauthor{gro96}~(\citeyear{gro96}, Fig. 1; plus sign $+$)
provide the distribution of Stokes $V$ \ion{Fe}{1} $\lambda$5250~\AA\ signals found
in the quiet Sun, being the average signal about $2 \times 10^{-3}$ (in units
of the continuum intensity). The authors also point out a calibration constant
from Stokes $V$ signal to flux density. It gives 2 G for the mean
observed signal. Since these signals fill 40\% of the solar surface,
the mean flux density is about 0.8 G. We assign to these data an
angular resolution twice
the sampling interval (some 2\farcs3).
Finally, \citeauthor{dom03}~(\citeyear{dom03}; diamond $\diamond$)
have recently found 17~G mean flux density 
in a 0\farcs 5 angular resolution magnetogram taken
with \ion{Fe}{1}~$\lambda$6302.5~\AA .

Figure \ref{inset} provides a different illustration of  the
similarities between the synthetic magnetogram
and the real Sun.  
It shows a real magnetogram of the quiet Sun
obtained in the  blue wing of \ion{Fe}{1} $\lambda$6302.5\ \AA~
(\citeNP{san00}, Fig. 1), i.e., using the observational setup 
that we have tried to reproduce (see \S 
\ref{calibration}). A patch of the real magnetogram with the size
of the numerical simulation has been replaced with the synthetic
one (Fig. \ref{inset} left). 
Except for the location of the inset, which we place in
an IN region, we have not used free parameters to
produce the combined magnetogram. The angular resolution has been chosen
to match the observation (1\arcsec), whereas the degree of polarization
comes directly from the synthesis.  Despite the absence of 
fine-tuning, it turns out to be extremely
difficult to distinguish the magnetogram with the artificial inset
(Fig. \ref{inset} left) from the real one (Fig. \ref{inset} right). 
In other words, the synthetic magnetogram contains structure with a spatial 
distribution similar to the observed one, and with a
degree of polarization which fits within the observed range.

\subsection{Asymmetries of the Stokes $V$ profiles\label{asym}}

In the absence of gradients of magnetic and velocity fields within 
the resolution element, the Stokes profiles have to obey well 
defined symmetries (see \S \ref{sec_introduction}). This
kind of symmetries
are never found in the quiet Sun,
instead profiles frequently show extreme asymmetries
(\citeNP{san96}; \citeNP{gro96}; \citeNP{sig99}).
Because the magnetic field in the numerical simulation has a highly intermittent
structure (\citeNP{cat99a}; \citeNP{emo01}), we expected the resulting
synthetic Stokes profiles to be asymmetric. 
In the following, we compare our synthetic profiles with observed ones
and study in a statistical sense their differences and similarities.
We proceed by first classifying 
the types of Stokes $V$  profiles produced by the simulation.
The $2.6\times 10^5$ Stokes $V$ profiles
in the snapshot were assorted using a cluster analysis algorithm identical
to that employed with real data by \citeauthor{san00} 
(\citeyear{san00}; \S 3.2). The procedure identifies and groups
profiles with similar shape, 
irrespective of their degree
of polarization and polarity, since it employs Stokes $V$ profiles
normalized to the largest blue peak (see \citeNP{san00} for
details).

The classification is summarized in Figure
\ref{vasym0}.
For each different class, the averaged Stokes $V$ 
profile over the ensemble is plotted. The mean is taken after each
individual profile has been multiplied by the
sign of its largest blue peak. This avoids cancellation of different
polarities. The mean profiles are numbered (upper left corner in each panel)
according to the percentage of profiles 
in the simulation that are elements of their class, with \#0 being the most
probable and 
\#17 the least probable. The probability to find a profile of a given
class is indicated in the upper right corner of each panel.
Note the various possibilities. From virtually anti-symmetric
Stokes $V$ (implying no asymmetry), to profiles with 
one (e.g., \#11 and \#13) or three lobes (e.g., \#12 and \#14).  All these asymmetries are produced by 
gradients of magnetic  and velocity fields along the \los .
The peak polarization is considerable,
some 1\% in units of the continuum intensity. This has to be the case to
yield the large flux density  shown in Figure \ref{mean}a for perfect 
angular resolution. Another detail worthwhile noting is the 
balance between the number of profiles having a large blue lobe and those
whose principal lobe is the red one.

The line shapes in Figure \ref{vasym0} are difficult to compare
with observed profiles since observations of the quiet Sun have
much lower angular resolution. 
Figure \ref{vasym1} shows  another 
classification having all the features explained above, except that
the simulation data have been smeared with a 1\arcsec\ seeing, typical
of the real observations. 
First, the polarization signals are reduced by one order of magnitude
with respect to the original profiles; compare them with those
in Figure \ref{vasym0}.
Second, new more complicated line shapes have arisen as the result
of the large spatial smearing (radiative transfer smoothes over only
100 km, whereas the spatial smearing does it over 725 $\times$ 
725 km$^2$). 
Stokes profiles having all these very asymmetric shapes are indeed
observed in the real quiet Sun (see \citeNP{san96}, Fig. 12;
\citeNP{sig99}, Fig. 7; \citeNP{san00}, Fig. 4, 5 and 6;
\citeNP{sig01}, Fig. 4).
This qualitative agreement is again a notable feature of the 
simulation, since there was no obvious a priori reason to expect 
it. 

However,
despite such general qualitative agreement,  one can find
quantitative differences between the synthetic and the observed
profiles. As it happens with the fully resolved profiles (Fig.~\ref{vasym0}),
the number of synthetic profiles 
having a principal blue lobe and those
with a principal red lobe is similar
(e.g., the pairs \#12 and \#18;
\#31 and \#34; \#33 and \#35). 
Such balance is not
present among the observed profiles, where the blue lobe 
usually  dominates (e.g., \citeNP{san00}, Fig. 4;
\citeNP{sig99}, Fig. 12). 
Another qualitative difference with observations concerns the 
profiles that are most frequently obtained in the simulation.
They show almost no asymmetry (see classes
\#0 to \#7). Contrariwise, the observational counterpart have 
a well defined asymmetry characterized by a large blue lobe 
(similar to those observed 
in plage and  enhanced network regions).
This lack of significant asymmetry can be traced back to the original
syntheses (Fig. \ref{vasym0}, classes \#0 to \#4), and therefore
to the variations along the \los\ of the magnetic field and velocity. 
The probable cause is discussed in the next paragraph.

Figure \ref{brms} summarizes the kind of variations along the \los\ 
existing in the numerical simulation. Several definitions
are required before it can be interpreted. 
The variations are described using standard statistical parameters, namely,
the  \los\ mean  
value $\overline{f}_{ij}$, the \los\ standard deviation $\overline{\delta f}_{ij}$, 
and the \los\ correlation coefficient $\overline{(fg)}_{ij}$,
\begin{displaymath}
\overline{f}_{ij}=n_z^{-1}\sum_{k=1}^{n_z} f_{ijk},
\end{displaymath}
\begin{equation}
\overline{\delta f}_{ij}=\Bigl[\overline{(f_{ijk}-\overline{f}_{ij})^2}\Bigr]^{1/2},
	\label{stat2}
\end{equation}
\begin{displaymath}
\overline{(fg)}_{ij}=
\bigl[\overline{(f_{ijk}-\overline{f}_{ij})(g_{ijk}-\overline{g}_{ij})}\bigr] \Big/
	\bigl[\overline{\delta f}_{ij}~\overline{\delta g}_{ij}\bigr].
\end{displaymath}
The arrays $f_{ijk}$ and $g_{ijk}$ may represent any component
of the fields, and their
indexes vary according to the position in the horizontal plane ($i$ and $j$)
and in the vertical direction ($k$).
The symbol $n_z$ stands for the number of grid points
in the vertical direction.
We should bear in mind that the \los\ averages defined in equations 
(\ref{stat2}) depend on the horizontal
coordinates.
We need to characterize the typical properties of these 
\los\ mean values
for each range of mean longitudinal magnetic field strength. 
For this purpose we define the average $\langle\overline{h}\rangle$ and 
the dispersion $\Delta\overline{h}$ of the quantity $\overline{h}$ among 
all those points in the simulation with
a given \los\ mean  longitudinal magnetic field $\xi$, i.e.,
\begin{equation}
\langle\overline{h}\rangle (\xi )=n_s^{-1}\sum_{i=1}^{n_x}\sum_{j=1}^{n_y}  \overline{h}_{ij}\ 
	P(\xi ,i,j),
\end{equation}
\begin{equation}
\Delta\overline{h}(\xi )=\Big[n_s^{-1}\sum_{i=1}^{n_x}\sum_{j=1}^{n_y}  
	\Big(\overline{h}_{ij}-\langle\overline{h}\rangle (\xi )\Big)^2\ P(\xi ,i,j)\Big]^{1/2},
\end{equation}
with
\begin{equation}
P(\xi ,i,j)=\cases{1, &if $\big|\xi-|\overline{B_z}|_{ij}\big| < \epsilon/2$, \cr
	0,& otherwise,}
\end{equation}
and 
\begin{equation}
n_s=\sum_{i=1}^{n_x}\sum_{j=1}^{n_y}  P(\xi ,i,j).  
\label{stat3}
\end{equation}
Note that the symbol $\overline{h}$ may represent  any of the
statistical parameters in equation (\ref{stat2}) ($\overline{f}$,
$\overline{\delta f}$ or $\overline{(fg)}$), and $n_x$ and $n_y$ correspond
to the grid points in the two horizontal directions.
The bin size $\epsilon$ has to be chosen to guarantee having enough points
per bin.
Following these definitions, we can now interpret 
Figure \ref{brms}. The different quantities represented are plotted as
function of the line-of-sight mean magnetic field.

The first panel in Figure~\ref{brms} shows the variation of
$\langle\overline{U_z}\rangle$ and $\langle\overline{\delta U_z}\rangle$.
The limited dispersion of the vertical velocities 
provides an explanation for the predominance of profiles with
small asymmetries. The dispersion must be of the order of the
line width to
produce a substantial modification of the line shape.
The typical standard deviation of the vertical velocities along the \los\
turns out to be between 0.2 and 0.3 km s$^{-1}$ 
($\langle\overline{\delta U_z}\rangle$, Fig. \ref{brms}a, the solid line),
whereas the line  widths  are of the order of 2 km s$^{-1}$ (see \S 
\ref{sec_spectra}). The magnetic field itself is probably not responsible for the 
moderate asymmetries since its variations along the \los\ are large. For example,
Figure \ref{brms}b shows the standard deviation of the
mean longitudinal magnetic field $\langle\overline{\delta B_z}\rangle$, which
is frequently larger
than the absolute  value of $\overline{B_z}$. 
In fact, the variations of the longitudinal magnetic fields  are so important that
$B_z$ very often changes sign along the \los\
(Fig. \ref{brms}c). Note that none of these statements on  
large magnetic field gradients
apply to the intrinsically strong fields, a case discussed
separately in the next paragraph. Above, say, 200 G,
the variations of field strength are mild
and the longitudinal field maintains a constant sign along
the \los\ (Figs. \ref{brms}b and \ref{brms}c).
Concerning the balance between the number of asymmetries towards the
blue and towards the red in Figures \ref{vasym0} and \ref{vasym1}, 
it is probably due to the lack of a 
definite sign for the correlation between magnetic field and 
velocity.
Works on the asymmetries in plage and network regions
repeatedly indicate the need for a negative correlation to account for the
observed preponderance of the Stokes $V$ blue lobe, explicitly,
\begin{equation}
\overline{(U_z |B_z|)} < 0,
	\label{correlation2}
\end{equation}
(see, e.g., 
	\citeNP{sol88}; 
	\citeNP{san88b}; 
	\citeNP{san89};
	\citeNP{gro88}; 
	\citeNP{gro89};
	\citeNP{san98a}). 
Such  condition is satisfied when downflows and magnetic fields 
are spatially separated, i.e., when
the strongest downflows tend 
to occur in weakly magnetized plasma.
(Note that $U_z > 0$ corresponds to downflows.)
One can see in Figure \ref{brms}d that 
the correlation between $|B_z|$ and $U_z$  has a well defined 
negative value only for the largest field strengths.
Since theses points 
represent a small fraction of the synthetic profiles,
they contribute very little to the classification in Figure \ref{vasym0}
which, consequently, shows no obvious preference for 
a blue asymmetry.
 
Those patches in the simulation with the largest 
field strength show a clear negative correlation fulfilling the
criterion in equation (\ref{correlation2}) (see Fig. \ref{brms}d).
Do they produce the observed Stokes $V$ profiles with a main blue lobe?
They do not, since the  asymmetry of the profiles emerging from 
concentrations of intense field strength are minimal.
This fact can be understood using Figures \ref{brms}a and \ref{brms}b,
which reveal gradients of both  magnetic field and velocity too small
for the requirements described in the previous paragraph.
Despite this apparent disagreement, the key ingredients to yield
the right shapes are already present in the simulation. 
If one artificially increases the gradients of magnetic field
and velocity already existing in the simulation, large asymmetries similar
to the observed ones automatically show up. 
Figure \ref{net_like} contains synthetic profiles emerging from the
three more intense 
magnetic concentrations in the snapshot (labeled as $a$, $b$ and
$c$ in Fig. \ref{magneto}, left). Note that they already have 
the Stokes $V$ asymmetries that characterizes network and IN regions.
(Figure~\ref{net_like}d includes one of these observed Stokes $V$ profiles
for reference, 
namely, one network profile in \citeNP{san00}).
In order to produce theses new synthetic profiles with enhanced asymmetries, 
we increased the intensity of the flow with respect to the scaling
in \S \ref{sec_spectra} by using  $f_u= 0.04$.
In addition, we increased the variations 
of magnetic field strength by averaging all the absorption and
emission over 0\farcs 5 of the simulation (i.e., 
over all the points in a box of this size).  The qualitative 
agreement between synthetic and real profiles is remarkable.

The above considerations lead to an interesting question. How should  the numerical simulations be modified
in order to produce asymmetries closer to the observed ones? 
First,  the dispersion of velocities at the smallest 
scales must be increased, which implies a decrease in the kinematic viscosity. 
Second, magnetic and non-magnetic regions should be even more intermittent
to strengthen the correlation (\ref{correlation2}). This can be achieved
by decreasing the magnetic diffusivity, which
would both increase the tangling of magnetic field lines and
 the dispersion of field strengths existing in the
intense concentrations.

\subsection{Broadening of the Intensity profiles\label{broadening}}

	One of the observational constraints on the
existence of a complex and tangled magnetic field in the solar photosphere
comes from the
work by \citeN{ste77} (see also \citeNP{unn59}; \citeNP{ste82}).
If a tangled magnetic field exist, it has to 
broaden the spectral lines of the solar spectrum
according to their magnetic sensitivities (i.e., according to their
effective Land\'e factors).
All other things being the same, those with
larger  sensitivity should be slightly broader. 
\citeN{ste77} looked for such effect in the solar unpolarized spectrum
with no success. From the
error budget of the measurement
the authors set
an upper limit to the field strength of the
existing fields (\citeNP{ste77}, equation [12]),
\begin{equation}
B_{\rm app}\leq 140\ {\rm G}.
	\label{limit}
\end{equation}
In this section we analyze if the magnetic fields in the
simulations produce line broadenings compatible with such 
observational upper limit. This exercise allows to address the
question of whether magnetic fields as intense 
as those in the simulation may exist on the Sun,
and still remain below the present observational detection limit.
We already know that the degree of circular polarization 
of the simulation stays
well within the observational bounds (\S \ref{flux_density}).
Here we address the
question from  a different perspective, using a totally different 
observational constrain that depends on the magnetic fields in a intrinsically
different way.

Our synthetic Stokes $I$ spectra have an excess of broadening
caused by the presence of magnetic fields. Figure
\ref{dif} shows the difference between the mean Stokes $I$
profile produced by the region, and the mean profile produced
when the syntheses are repeated with no magnetic field, but keeping
everything else identical.
The magnetic profile is broader and shallower at the line center,
producing a residual with three lobes.
Is this extra broadening compatible with the 
observational limit (\ref{limit})?
A detailed modeling of the procedure
employed by \citeN{ste77} is clearly beyond our possibilities,
since it requires
the synthesis of hundreds of spectral lines
with different temperature and magnetic field sensitivities.
Fortunately, the essence of the procedure is simple.
If two lines are
identical except for their magnetic sensitivity, the small excess of width
$\Delta w$
can be directly related to an {\em apparent} magnetic field strength $B_{\rm app}$
according to the rule,
\begin{equation}
B_{\rm app}=\delta \sqrt{(\Delta w/w)}.
	\label{broa3}
\end{equation}
The scale factor $\delta$ depends of the wavelength,
the difference of Land\'e factor, and the mean line width $w$.
For two lines with the wavelength and strength of
\ion{Fe}{1} $\lambda$6302.5~\AA , one magnetic and
another one non-magnetic, 
the scaling factor turns out to be
\begin{equation}
\delta\simeq 2.1 \times 10^{3}\ {\rm G},
	\label{broa4}
\end{equation}
which follows from the equations
(7) and (8) in \citeN{ste77}.
The estimate of the apparent magnetic field strength using
equations (\ref{broa3}) and (\ref{broa4}) is
a matter of determining the excess of broadening $\Delta w/w$
associated with the residuals in Figure \ref{dif}.
Following \citeN{ste77}, the two
mean intensity profiles where fitted using Gaussian functions, 
i.e., 
\begin{equation}
I_c - I \propto \exp\big[-(\lambda/w)^2\big],
\end{equation}
where $\lambda$ stands for the wavelength relative to the
line center and $I_c$ represents the continuum intensity.
Then the difference between the widths $w$ for the syntheses 
with and without 
magnetic field directly yields
\begin{equation}
\Delta w/w\simeq 1.1 \times 10^{-2},
\end{equation}
or, using equations  (\ref{broa3}) and (\ref{broa4}),  
\begin{equation}
B_{\rm app}\simeq 220\ {\rm G}.
\label{limit2}
\end{equation}
Note that the Gaussian fits reproduce fairly well
the difference between the two synthetic profiles (see the dashed line
in Fig. \ref{dif}).

The field strength
of our synthetic profiles (equation [\ref{limit2}]) apparently 
come into conflict with the observational limit in equation (\ref{limit}).
Should the inconsistency be real, it points out 
an excess of magnetic fields in the numerical
simulations as compared to the solar case (an excess of magnetic field strength,
area coverage of the fields, etc.).
However, the marginal discrepancy is probably not significant
in view of the uncertainties affecting
both the syntheses and the observational limit.
Actually,  
the similarity between  the observational limit and
the predicted  width should be understood
as real chance to test
whether complex tangled fields like those in the
numerical simulations are present in the solar
photosphere. A slight refinement of the currently available
diagnostic tools (e.g., Stenflo \& Lindegren's technique)
should be able unambiguously to confirm them or discard  them.
We return to this point  in \S \ref{conclusions}.

\section{Hanle signals produced by the dynamo simulations\label{hanle}}

	Up to now we only consider polarization signals generated
by Zeeman effect. However, 
solar turbulent magnetic fields have been inferred using Hanle effect
signals\footnote{The Hanle effect is a purely non-LTE phenomenon  whose
details and subtleties are still a subject of active research. Roughly 
speaking, the magnetic field modifies the polarization of the
light that we detect after scattering 
(see, e.g., \citeNP{lan92}; \citeNP{ste94}).}. 
They indicate the presence of turbulent magnetic fields in the
upper photosphere-lower chromosphere
with a field strength between 5 and 60 G (e.g., \citeNP{fau95};
\citeNP{bia99}).
These strengths may seem too low compared to those assumed
in  this work,  with a mean value larger than
100 G (e.g., \S \ref{broadening}). Therefore, we felt compelled to 
estimate the Hanle signals expected from the dynamo
simulation and compare them with observed values.
Once more we face the question of whether magnetic fields similar
to those in the simulations may exist and still produce
observable effects within solar values.

A complete
Hanle effect synthesis similar to that carried out for the Zeeman effect
is clearly beyond the scope of this paper. Fortunately, one can 
estimate the level of Hanle signals for one of the typical lines
used in Hanle effect based diagnostics.  Considering the
Sr~{\sc i} line at $\lambda$4607~\AA , the depolarization produced by
a magnetic field  can be expressed  as 
\begin{equation}
Q/Q_0\simeq W_B=1-{2\over 5}\Big({{\gamma_H^2}\over{1+\gamma_H^2}}+
	{{4\gamma_H^2}\over{1+4 \gamma_H^2}}\Big),
	\label{hanle1}
\end{equation}
where 
$Q/Q_0$ is the ratio between the observed linear polarization
$Q$ and the polarization  expected if there were no magnetic field $Q_0$. 
The
symbol $\gamma_H$ parameterizes the magnetic field strength of the
turbulent field $B$,
\begin{equation}
\gamma_H = B/B_H.
	\label{hanle2}
\end{equation}
The normalization factor  $B_H$ scales linearly 
with the radiative transition rate plus the 
depolarizing collision rate.
The relationship (\ref{hanle1}) is an approximation
which works  well for this particular line in standard
quiet Sun model  atmospheres (see \citeNP{fau01}, where
one can also find the dependence 
of $B_H$ on the density and temperature of the atmosphere).
For additional details see, e.g., \citeN{lan85b} and
\citeN{tru99b}.

In case that the magnetic field strength is not unique but
has a distribution of values, the mean signal $\widetilde{Q}/Q_0$ 
turns out to be equal 
to the mean depolarization factor considering the distribution 
of the field strengths
$\widetilde{W}_B$ (see \citeNP{lan85b}, \S 3).
Note that $W_B$ no longer varies with $B$ as $B \gg  B_H$ ($\gamma_H \gg 
1$),
therefore the average depolarization $\widetilde{W}_B$ bears
no information 
on the large field strengths that
may exist in the distribution.
In other words, the mean signal will always be biased
towards weak field strengths (e.g., \citeNP{tru01}, Fig. 3).
This fact
could reconcile the large fields in 
the dynamo simulations with the observed Hanle depolarization
signals. In order to check such possibility,
we evaluate the mean $\widetilde{W}_B$ for the snapshot 
studied in the paper.
The transition and collision rates  required to compute $B_H$
are determined, according to the prescription in \citeauthor{fau01}(\citeyear{fau01}, \S 3), 
using the quiet Sun densities and temperatures given by \citeN{mal86}.
We only consider the photospheric layers. The expected signals, including 
all the magnetic fields in the snapshot describing the distribution of $B$, 
are represented in Figure \ref{hanle_fig}. The depolarization changes with height due to
the variation of temperature and density in the model atmosphere, which
modify $B_H$. Figure \ref{hanle_fig} includes  the 
range of Hanle signals observed close to the disk center,
namely,  $\widetilde{Q}/Q_0=0.50 \pm 0.18$
(quoted by \citeNP{fau95}, from  the measurements of 
\citeNP{ste82} at an heliocentric angle whose cosine is  $\mu=0.81$). 
The observational signal has been chosen to be as close as possible to the
disk center to sample  deep photospheric layers. The measurement of depolarization 
is ascribed by \citeN{fau95} to a range of heights between  200 km and 400 km.
Although these layers
are still too high for the range of heights that we assign to the
simulation in the Zeeman syntheses (the base of the photosphere),
the  depolarization turns out to be within the 
bounds set by observations (see Fig. \ref{hanle_fig}, the solid line).
The agreement improves taking into account that the turbulent field strengths
are expected to decrease with height in the atmosphere 
(e.g. \citeNP{fau95}). The dashed line in Figure \ref{hanle_fig} 
has been computed assuming the field strengths to be 
a factor of two smaller than those in the original snapshot.
In short, our simplified estimate 
indicates no obvious
inconsistency between the simulations 
and the observed Hanle effect depolarization signals for
\ion{Sr}{1}~$\lambda$4607~\AA .


\section{What spatial resolution is necessary to resolve 
the numerical data?\label{telescope}}

Answering this question 
may be relevant to design  of 
the next generation solar telescopes (Rosner \& Beckers, 2001,
private communication).
For the sake of simplicity, 
we adopt the magnetic flux 
as the physical parameter to be determined. 
It is routinely measured using standard
instrumentation, and it suffers a severe bias due to insufficient
spatial  resolution (\S \ref{flux_density}).
The fraction of  magnetic 
flux in the simulation that one still detects in synthetic 
magnetograms is used 
to quantify the required angular resolution. 
This information is contained in Figure \ref{mean}b, which we 
plot in a slightly different way in Figure \ref{teles_diam}. 
Abscissae and ordinates have been interchanged, and the angular
resolution is also  presented in terms of the diameter of 
the ideal telescope achieving the angular resolution (see
equation [\ref{eq1}]).This representation suits our present
purposes.
The curve is shown for two different sensitivities, i.e.,
when (virtually) all polarization signals 
are detected, and when only those larger than 20 G
are above the noise level. The following conclusions follow  from Figure \ref{teles_diam}:
\begin{enumerate}
\item Irrespective of the resolution, 
	one only detects 80~\% of the existing flux.  The  residual
20~\% cancellation is due to the smearing along the \los\
 produced by the radiative transfer. 
Surpassing this upper limit is not a question of increasing the 
telescope size but rather it demands  improving the diagnostic techniques
used to measure the magnetic flux.
\item Detecting 50~\% of the flux would require an angular resolution of 
110~km or a telescope of 85~cm. 
Detecting another 20~\% additional  magnetic flux demands  
resolving 40~km and so a telescope some three times larger 
(2.3~m).
\item Points~\#1 and \#2 refer to observations without noise. If one
considers a  detection threshold of 20~G (which corresponds to a
degree of polarization of some 0.3~\%, according to the calibration
in \S \ref{calibration}), then only 70~\% of the flux 
present in the simulation
can be detected. 
On the 
other hand detecting 50~\% implies an angular resolution of 70~km or a diameter
of $\simeq$~1.3~m. 
\item The slope of the solid line in Figure \ref{teles_diam} drastically changes when
trying to detect more than, say, 65~\% of the original flux. Going
beyond this point requires a large increase of telescope
diameter for a limited increase of additional signal. This 65~\% 
may represent an  optimal compromise between 
resolution and telescope  size, and
it corresponds to
60~km on the Sun  or a diameter of  1.5~m.
\end{enumerate}

\section{Discussion\label{conclusions}}

The dynamo simulation by \citeN{cat99a} (see also \citeNP{emo01})
produce magnetic fields whose structure resembles in many ways 
the quiet Sun magnetic fields (\S \ref{sec_introduction}).
Although the simulations were not designed for a  
realistic description of the solar conditions, the quantitative
comparison with the quiet Sun that we undertake
is useful for a  variety  of
reasons. It allows to judge whether, and to what extent, the turbulent
dynamo provides
a paradigm to describe the quiet Sun magnetism. It allows
to identify, study and understand the difficulties and biases faced by
the current observational techniques when applied to very complex fields.
It helps identifying physical ingredients that are missing in the simulations, an 
exercise helpful to guide future  numerical work. 
Finally, it may suggest the type of technical 
developments needed to measure the properties of a magnetic field 
with the complexity present in the numerical data.
Keeping in mind all these reasons, we synthesize the polarization
emerging from the simulation. Specifically, we choose a snapshot
of the time series representative of the stationary regime, which
renders some 2.6 $\times$ 10$^5$ individual spectra. The
assumptions and limitations of our Milne-Eddington approach to the
synthesis are discussed  in \S \ref{sec_spectra}. They let us calculate the 
polarization produced by a magnetically sensitive spectral line similar to 
\ion{Fe}{1} $\lambda$6302.5~\AA . The synthetic spectra have been analyzed
in the light of observations of quiet Sun magnetic fields.
The comparison reveals similarities and differences. In addition, it
provides some hints and caveats to keep in mind when interpreting observations.
These three aspects of the synthetic spectra are discussed next.

\begin{figure}
\epsscale{.8}
\plotone{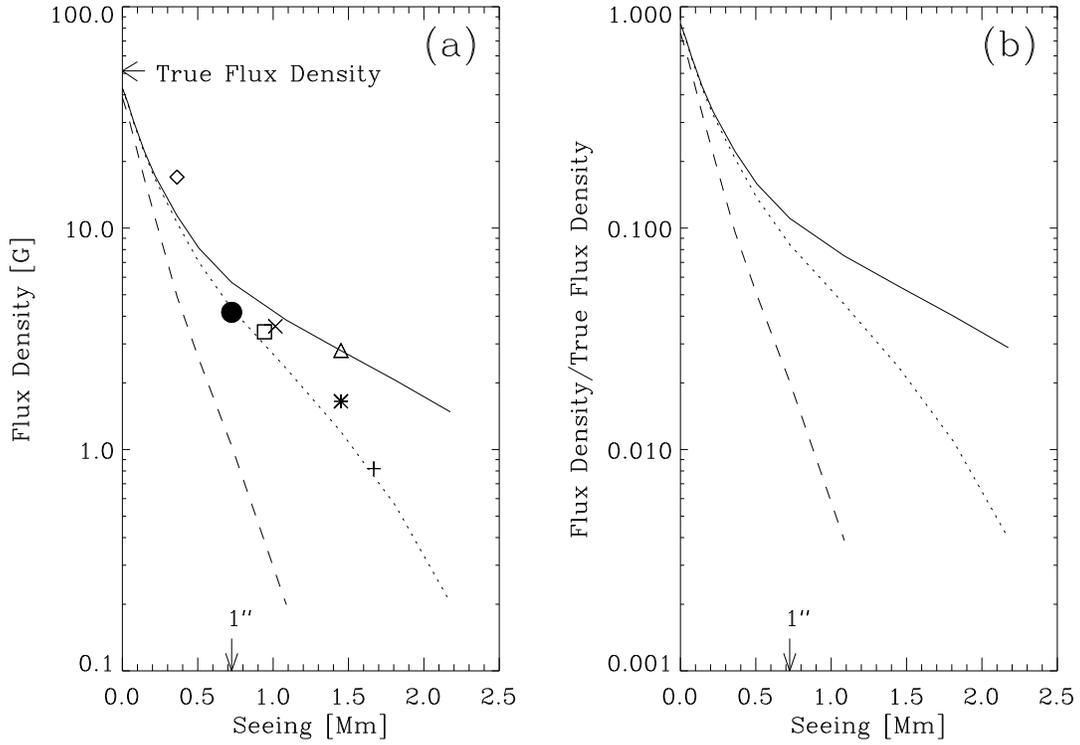}%
\caption{Mean flux density as a function of the spatial  resolution
of an hypothetical observation. The three different curves represent different
sensitivities of the observation (0.5 G, the solid lines;
5 G, the dotted lines; 20 G, the dashed  lines). (a) The flux density in
the synthetic magnetogram is given
in G. The symbols represent real observations of IN
fields. For the equivalence between symbol and 
reference, see \S \ref{flux_density}.
(b) The flux density has been normalized to the true mean flux density,
i.e., the mean value of $|B_z|$ in the simulation. It helps understanding
the large fraction of missing flux for the sensitivity and angular resolution
of  a typical observation (say, 5 G and  1\arcsec).
Note the 20 \% reduction for perfect angular resolution, which is produced
by the   radiative transfer smearing along the line-of-sight.
The arrow on the abscissae points out 1\arcsec .
}
\label{mean}
\end{figure}

\begin{figure}
\epsscale{.6}
\plotone{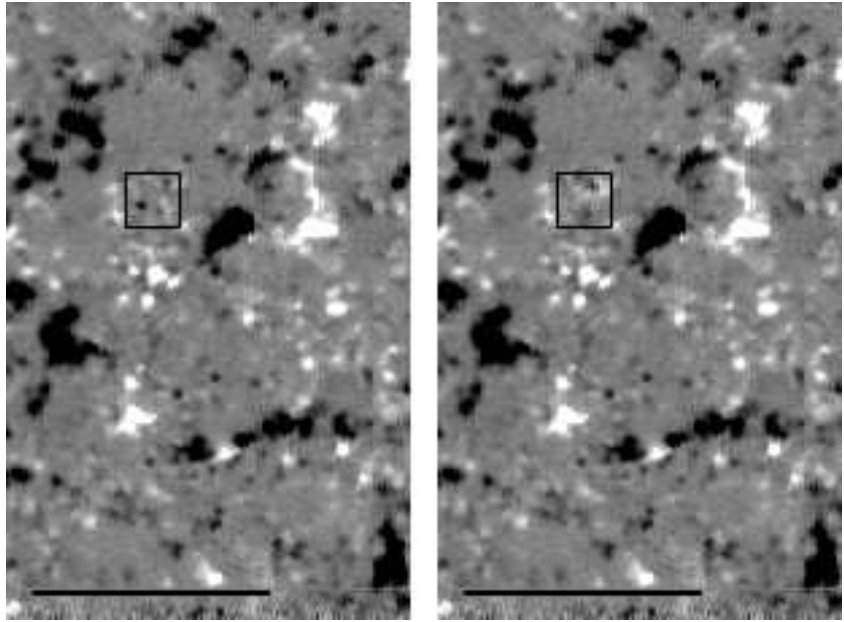}
\caption{Synthetic magnetogram embedded in the real magnetogram
	of a quiet Sun region: inset within a box
	in the left image. The right image shows the original
	magnetogram, including the box for reference.
	The bar in the bottom corresponds to 25 000 km on the
	Sun, i.e., the linear dimension of a typical network cell.
	Note how the synthetic magnetogram fits in smoothly within
	the real magnetogram. It produces the right polarization
	for IN fields and it also has 
	the proper spatial scales.
	}
\label{inset}
\end{figure}

\begin{figure}
\epsscale{.8}
\plotone{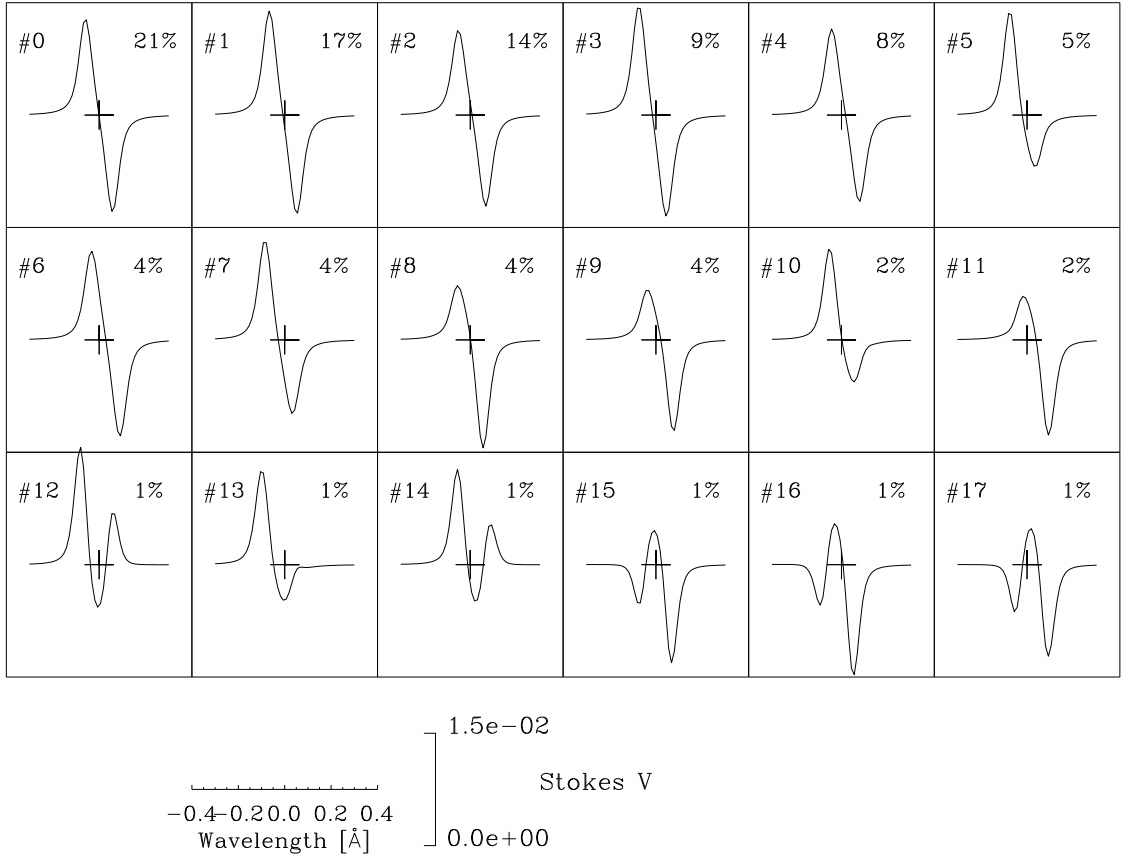}
\caption{Types of Stokes $V$ profiles produced by the numerical simulation.
All plots share a common scale which is indicated at the bottom of the
figure.
The label of each profile includes a number 
for cross-reference, and
the percentage of profiles in the simulation belonging to the category.
The profiles have been taken from the original simulation,
with no spatial smearing. The large plus sign of each plot indicates the origin
of the horizontal and vertical scales, shown at the bottom of the figure.
The Stokes $V$ profiles are normalized to the continuum intensity.
}
\label{vasym0}
\end{figure}

\begin{figure}
\epsscale{.8}
\plotone{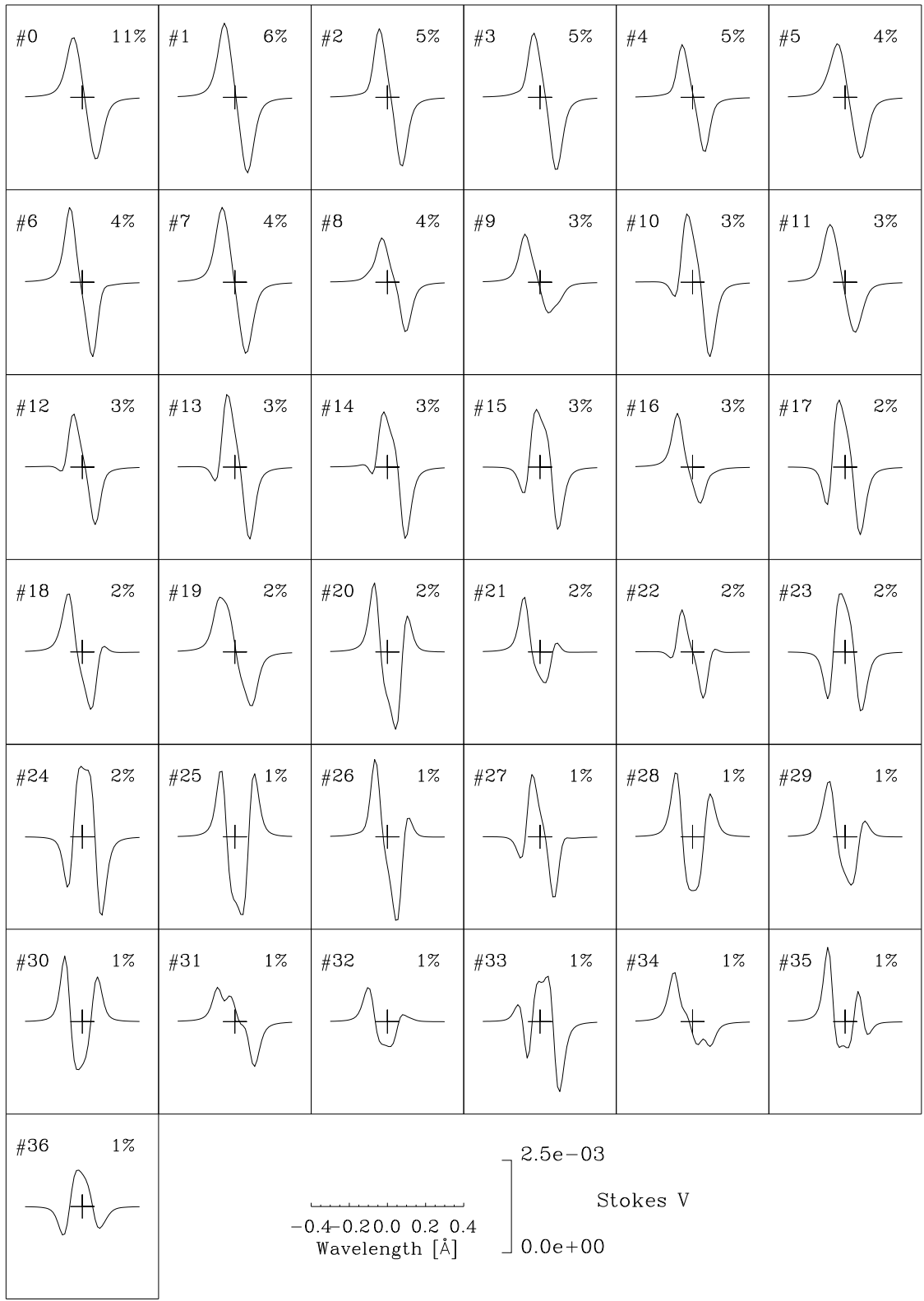}
\caption{Same as Figure \ref{vasym0}, except that the synthetic
profiles have been  smeared with a 1\arcsec\ seeing. The variety
of possibilities has  increased.
}
\label{vasym1}
\end{figure}

The magnetograms produced by the numerical data
are in good agreement with those observed in the Sun (\S \ref{flux_density},
Figs. \ref{mean}a and \ref{inset}). The agreement, however is only achieved 
after a severe canceling of signals produced by the  poor
spatial resolution of 
the current observations. More than 90 \% of the unsigned magnetic
flux existing in the numerical data does not appear in
magnetograms with 1\arcsec angular resolution (Fig. \ref{mean}b). 
In other words, it is this 2 to 10 \% residual that agrees with
the observed signals. The Stokes $V$ profiles emerging from the simulation are
frequently very asymmetric, in qualitative agreement with
observations (see references in \S \ref{sec_introduction}).
In particular, Stokes $V$ profiles with three and more
lobes result from the existence of two opposite polarities in 
the resolution element. Part of these asymmetries are produced by 
gradients of magnetic field and velocity along the \los\
(Fig. \ref{vasym0}), but the Stokes $V$ profiles also owe 
much of their shapes to variations across the \los\
within the resolution element (cf. Figs. \ref{vasym0} and
\ref{vasym1}). We studied in \S \ref{broadening}
the excess of broadening of the synthetic intensity profiles due to the
presence of a magnetic field. The additional broadening
is found to be close to, but typically in excess of, the upper limit of 140 G
set by \citeN{ste77}.  The uncertainties of the measurement
could easily explain this discrepancy,
therefore,  we understand this marginal disagreement 
as an invitation to
revisit the  work by \citeN{ste77} and improve the
sensitivity by a factor two. This should be enough clearly to
confirm or discard a solar magnetic field with the features present
in the numerical data. So far only Zeeman signals 
of typical photospheric lines have been mentioned.
We also estimate in \S \ref{hanle} the Hanle depolarization to be 
expected if the simulation is placed at various heights
in the photosphere. All the uncertainties notwithstanding, the
depolarization signals for \ion{Sr}{1}~$\lambda$4607~\AA\  turn out to be
within the observed bounds.

A quantitative analysis of the Stokes $V$ asymmetries reveals 
real discrepancies between the synthetic spectra  and
the quiet Sun. Most of the synthetic profiles show mild asymmetries,
well below the mean observed values. Moreover, the synthetic
spectra do not contain the clear observed tendency for the Stokes $V$ blue
lobe to dominate. Such trend is present even in the largest polarization
signals, which trace big concentrations
of magnetic field. We believe that the cause of the discrepancy
is twofold.
First, the dispersion of velocities between nearby pixels is too small.
Second, the spatial separation between strongly magnetized and unmagnetized
plasmas is too large. These two 
factors minimize the asymmetries 
in the intergranular lanes, despite the fact that
the kind of correlation between 
magnetic field and velocity already existing in the simulation 
produces the right asymmetry.
Arguments for such explanation are given in \S \ref{asym}. 
As a support of this view, we artificially increase the dispersion
of velocities and magnetic fields existing in the simulation to synthesize
spectra in three particularly large magnetic concentrations.
The resulting Stokes $V$ profiles look very much like those observed in the quiet Sun 
network
(synthetic and observed profiles are represented in Fig. \ref{net_like}).
If we have correctly identified
the origin of the differences,
what must be modified in the numerical simulation
in order to produce asymmetries closer to the observed ones? One
would need to increase the dispersion of velocities  and 
magnetic fields at the smallest spatial scales, keeping the relationship 
between them already existing in the simulation. 
This could be attained  by increasing the
Reynolds numbers of the simulations.

There is another discrepancy between the synthetic and  
the observed profiles which has not been mentioned yet. The analysis of Stokes 
$V$ shapes observed in the quiet Sun reveals that kG field strengths are common 
(see \citeNP{san00}, \citeNP{soc02}). 
There are also weaker fields (e.g., \citeNP{lin99}; \citeNP{col01}), but
the frequency of kG is certainly larger than that present in the simulation.
This difference can be readily pin down
to the incompressibility of the simulation (\S \ref{sec_mhd}), which hampers
the strong evacuation of the plasma, an ingredient
always associated with the existence of kG fields in the
photosphere.

\begin{figure}
\epsscale{1.0}
\plotone{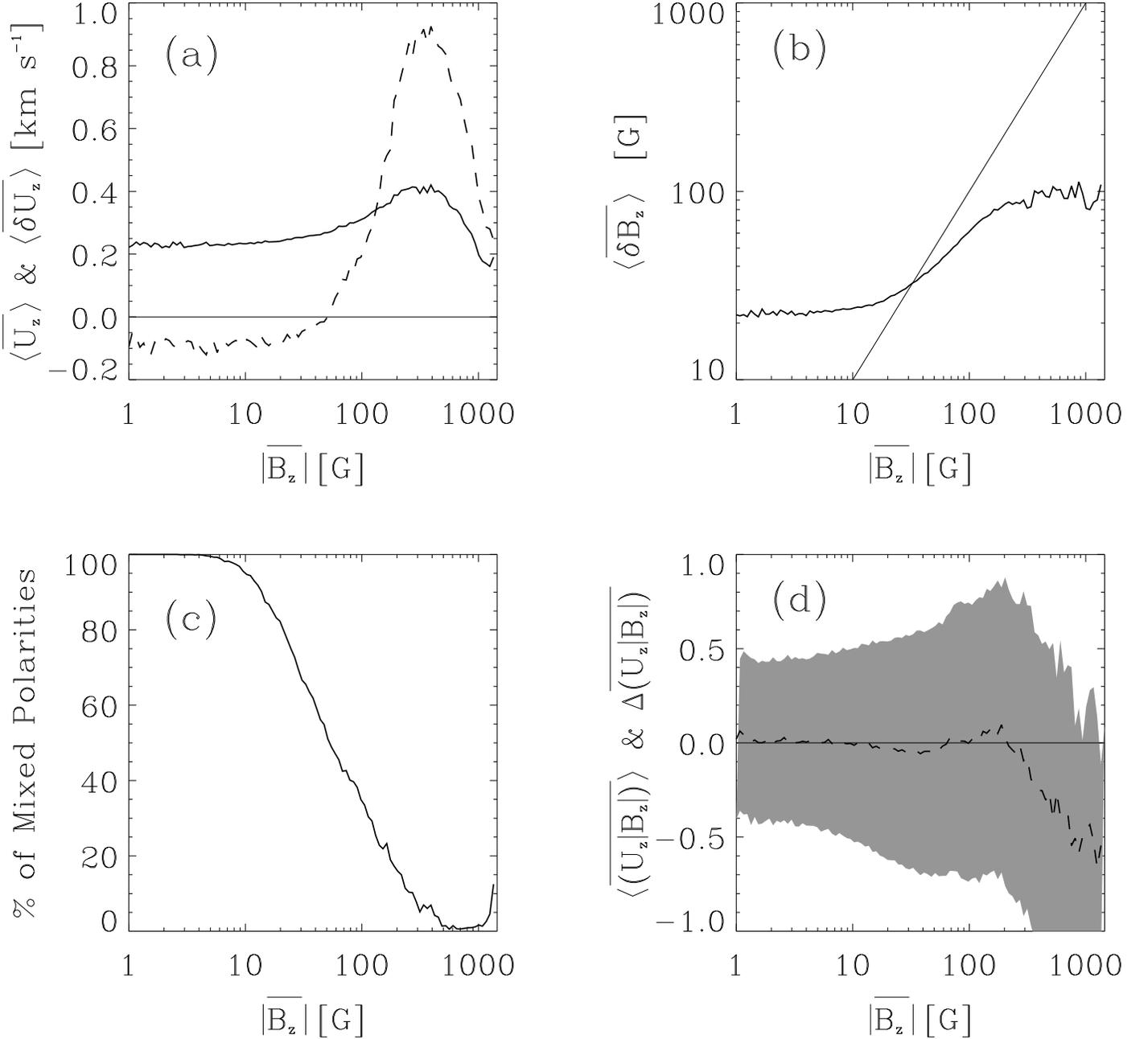}
\caption{Variations along the line-of-sight of the longitudinal
magnetic field and the longitudinal velocity. 
We represent mean values among all those points
in the simulation having the \los\ mean longitudinal magnetic field given
in the abscissae ($\overline{B_z}$; see equations [\ref{stat2}] to
[\ref{stat3}] for definitions). 
(a) Mean vertical velocity (the dashed line) as well as the
mean standard deviation among the velocity fluctuations along the line-of-sight
(the solid line).
($U_z > 0$ corresponds to downflows or redshifts.)
(b) Mean standard deviation of the magnetic field fluctuations
along the line-of-sight. The oblique straight line $y=x$ is included
for reference.
(c) Percentage of points where the longitudinal magnetic field
changes sign along the line-of-sight.
(d) Correlation coefficient for the fluctuations of magnetic field
and velocity along the line-of-sight. The dashed line corresponds to the mean
value, $\langle\overline{(U_z|B_z|)}\rangle$, whereas the shaded region shows one 
standard deviation about this mean value, $\langle\overline{(U_z|B_z|)}\rangle
	\pm\Delta\overline{(U_z|B_z|)}$.
Velocities and magnetic fields are in km s$^{-1}$ and G, respectively.
The correlation coefficients are dimensionless.
}
\label{brms}
\end{figure}

\begin{figure}
\epsscale{.6}
\plotone{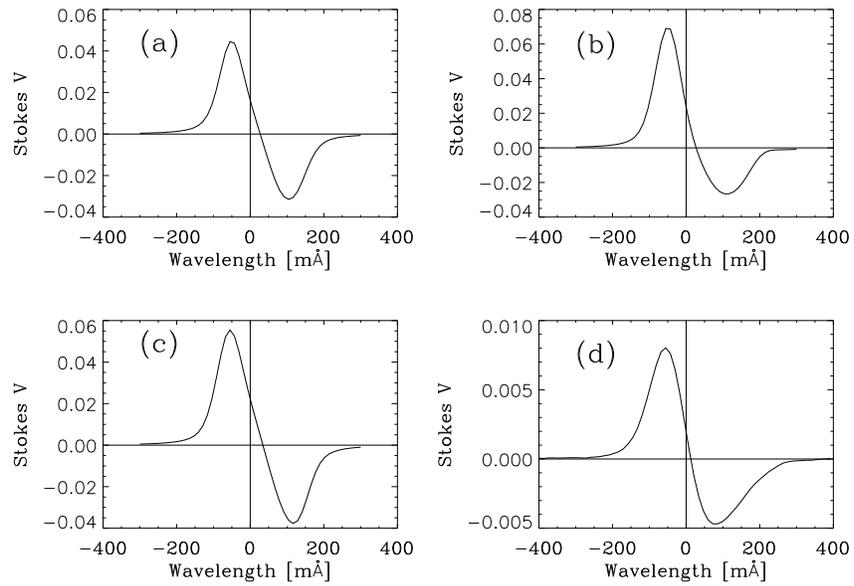}
\caption{(a), (b), and (c) correspond to the synthetic Stokes $V$ 
profiles resulting form the
average of the patches $a$, $b$ and $c$ in Figure \ref{magneto}.
(d) Observed profile shown for reference. It is the 
class 3 profile in Figure 4 of \citeN{san00}.
}
\label{net_like}
\end{figure}

\begin{figure}
\epsscale{0.8}
\plotone{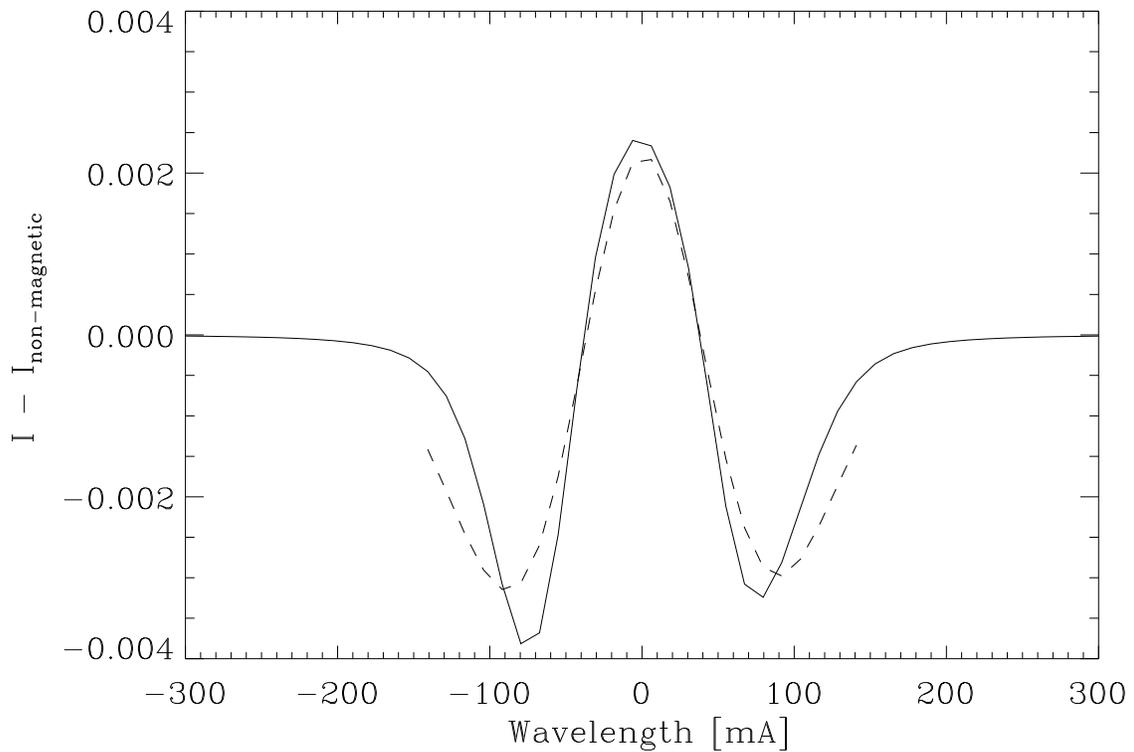}
\caption{Difference between the Stokes $I$ profiles synthesized
with and without magnetic fields ($ I - I_{\rm non-magnetic}$).
The difference has been modeled as the difference of two Gaussian
functions (the dashed line) whose different widths correspond to
an apparent magnetic field of the order of
220 G. (See text for details.)
}
\epsscale{1.0}
\label{dif}
\end{figure}

\begin{figure}
\epsscale{0.8}
\plotone{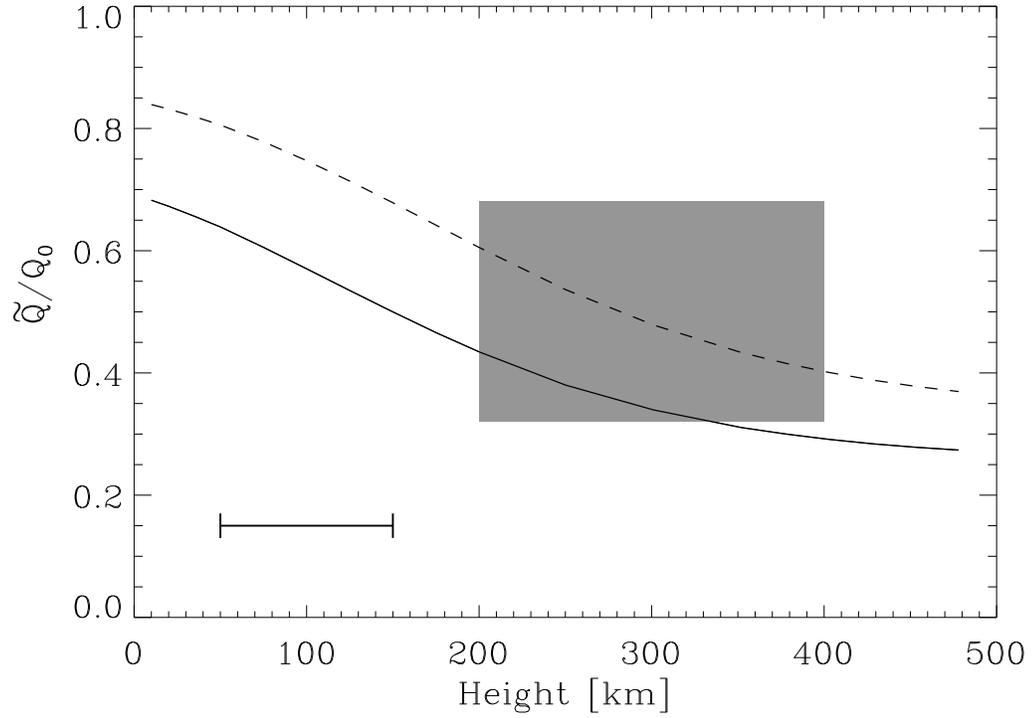}
\caption{Hanle depolarization 
produced by the dynamo simulations. The depolarization
of Sr {\sc i} $\lambda$4607 \AA\ is represented
versus the height in the atmosphere where the depolarizing
collisions are evaluated.  The two types of line represent
different scalings of the dimensionless magnetic field
strength: the one used along the paper (the solid line),
and a factor two smaller (the dashed line).
The shaded region corresponds to measurements of the
depolarization close to the solar disk center (see \citeNP{fau95}).
The segment points out the range of heights that we ascribe to the
dynamo simulation for the synthesis of \ion{Fe}{1}~6302.5~\AA .
}
\label{hanle_fig}
\end{figure}

\begin{figure}
\epsscale{0.8}
\plotone{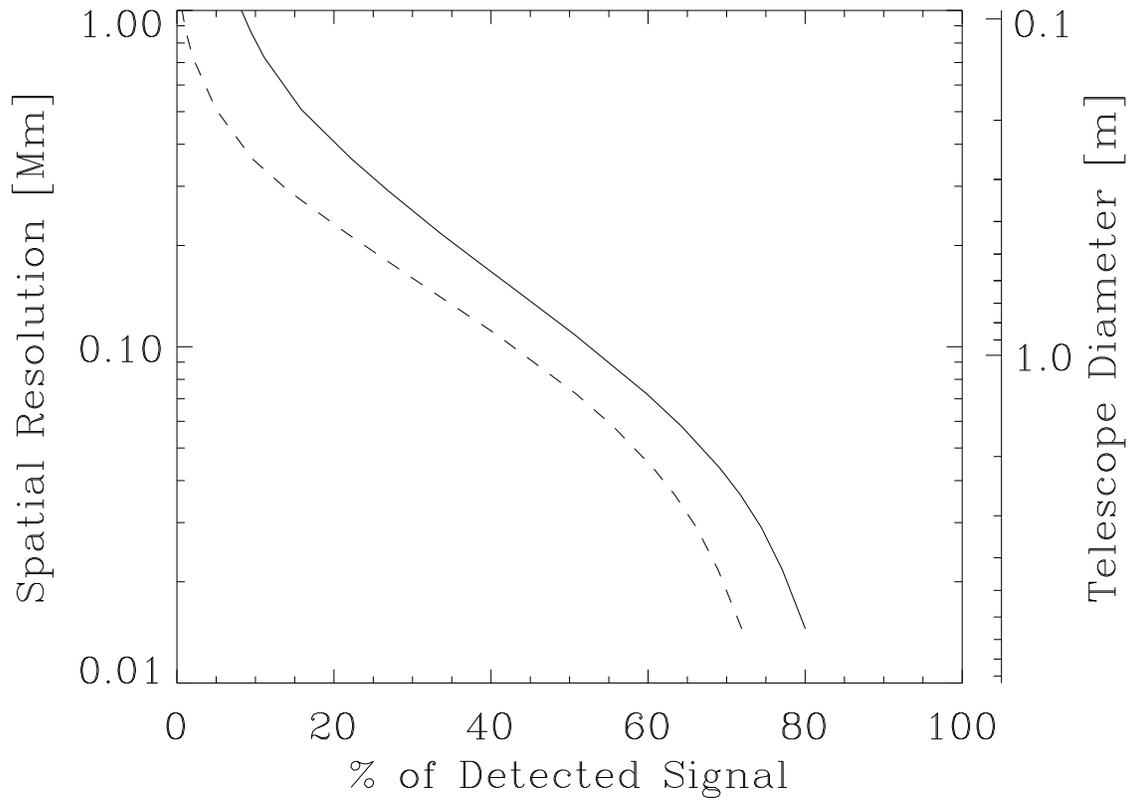}
\caption{Spatial resolution required to detect a given
	fraction of the magnetic flux density present in 
	the numerical simulation. The diameter of a diffraction
	limited telescope that yields the required spatial resolution
	is given in the second axis of ordinates. The solid line represents 
	observations without noise whereas the dashed line
	corresponds to a noise of 20 G, equivalent to a degree of polarization
	of some 0.3\% . Angular resolutions are in
	Mm and telescope diameters in m.}
\label{teles_diam}
\end{figure}

We have found several similarities  between the synthetic
spectra emerging from the dynamo simulation and 
observations of quiet Sun magnetic fields. Such  (sometimes unexpected)  
agreement seems to point out that the simulation  already
capture some of the essential features characterizing the quiet Sun magnetic fields. 
These arguments enables us to proceed by analogy and 
propose the existence of magnetic fields on the Sun
similar to  those in the numerical simulations. 
Several interesting consequences follow from this assumption:
\begin{enumerate}
\item The amount of flux inferred by conventional
      magnetograms  depends very much on  the angular resolution and the
      sensitivity of the measurements. Good or very good sensitivity is mandatory
      to detect any fluxes with the canonical 1\arcsec angular resolution. For
      example, the structures in the numerical simulation would
      be very difficult to detect in magnetograms like those provided
      by the MDI instrument on board the SOHO spacecraft, with a sensitivity of some
      15 G (e.g., \citeNP{liu01}). According to Figure \ref{mean}b, only
      1\% of the original magnetic flux is detectable at this sensitivity level.
      On the other hand, the sensitivity is not so critical upon improvement
      of the angular resolution. In the  limit of perfect angular resolution the
      signals become of the order of 50 G, corresponding to
      a degree of circular polarization of 1\% (see Fig. \ref{vasym0}).

\item The amount of unsigned flux existing in the simulation 
   is also very large in absolute terms. Assuming that the full solar surface 
   were covered by magnetic fields like those in the numerical simulations, the associated
   total unsigned flux would be of the order of
   $3.1~\times~10^{24}$~Mx. This figure is a factor 4.5 times larger
   than the total flux detected at solar maximum with conventional 
   techniques (\citeNP{rab91}, Fig. 5). Should such large amounts of
   hidden flux exist in the photosphere, it would have to 
   interact with the other classical  manifestations of the
   solar magnetism (active
   regions, solar cycle, coronal heating, etc).
   This point in particular deserves further investigations.

\item We discuss in \S \ref{telescope} the  telescope size required to detect the
	magnetic structures observed in the simulations. This is a preliminary estimate
	 which certainly has to be revised and
	refined. (For example, we do not address the question of whether 
	the 60 km optimum spatial resolution is set by the photon mean-free-path
	or the diffusion processes included in the MHD simulation.)
	However one conclusion of this preliminary study stands out.
	A sizeable fraction of the magnetic flux is not detected due to
	the radiative transfer smearing along the \los . Improving such bias 
	is not a matter  of increasing the telescope size and polarimetric 
	sensitivity. It only depends on  interpreting the observed polarization
	using techniques that account for the \los\ smearing.
	Keeping in mind the complexity of the fields (e.g., the longitudinal
	magnetic field almost always changes sign along the \los ; see Fig.
	\ref{brms}c), the use of inversion techniques that assume optically
	thin fluctuations of the magnetic field seems to be a reliable
	selection (e.g., the MISMA inversion code in \citeNP{san97b}).

\item The Stokes asymmetries contain information on the magnetic structuring
	at spatial scales smaller than the resolution element, an information
	unique and valuable (see \S \ref{sec_introduction}).
	The syntheses show that both  variations along the \los\ and 
	across the \los\ are responsible for the asymmetries. Consequently,
	the inversion techniques aimed at understanding  the origin  of the
	asymmetries, and therefore at measuring the properties of the photospheric
	magnetic fields, have to incorporate the two aspects of the
	spatial variations.
\end{enumerate}
The importance of the  above conclusions depends to some extent, on the degree of realism of 
the simulations. However, the tendencies 
that they imply are probably correct; e.g.,  the  existence of
large amounts of solar magnetic flux missing in conventional measurements,  or
the need for non-standard inversion techniques to determine the properties
of the  quiet Sun magnetic fields. These problems are important
enough to deserve a close follow-up.

\acknowledgements

	The interest to synthesize the spectra of turbulent dynamo
	simulators arose from conversations with T. Bogdan back
	in 1998.
	Both R. Rosner and J. Beckers pointed out the importance
	of estimating the telescope size required to spatially resolve
	the simulation.
	Thanks are due to F. Kneer, F. Stolpe, and M. Collados for
	providing the flux densities cited in \S \ref{flux_density},
	and to an anonymous referee for helpful comments.
	Thanks are also due to J. Stenflo for clarifications 
	on the Hanle effect based measurements of the
	turbulent magnetic field strength. 
	Discussions with R. Manso and J. Trujillo Bueno were extremely
	helpful to elaborate \S \ref{hanle}.
	This work has been partly funded by the Spanish Ministry of
		Science and  Technology, under projects AYA2001-1649
		and DGES 95-0028-C.
	Two of the authors FC and TE were partially supported by 
	NASA grant NAG5-10831 at the university of Chicago.

\def\jfm{J. Fluid Mech.}

\newpage

%

\end{document}